\documentstyle[12pt]{article}

\renewcommand{\theequation}{\thesection.\arabic{equation}}

\newlength{\extraspace}
\newlength{\extraspaces}
\newcounter{dummy}

\newcommand{\be}{\begin{equation}
\addtolength{\abovedisplayskip}{\extraspaces}
\addtolength{\belowdisplayskip}{\extraspaces}
\addtolength{\abovedisplayshortskip}{\extraspace}
\addtolength{\belowdisplayshortskip}{\extraspace}}
\newcommand{\ee}{\end{equation}}

\newcommand{\ba}{\begin{eqnarray}
\addtolength{\abovedisplayskip}{\extraspaces}
\addtolength{\belowdisplayskip}{\extraspaces}
\addtolength{\abovedisplayshortskip}{\extraspace}
\addtolength{\belowdisplayshortskip}{\extraspace}}
\newcommand{\ea}{\end{eqnarray}}

\newcommand{\baa}{
\addtocounter{equation}{1}
\setcounter{dummy}{\value{equation}}
\setcounter{equation}{0}
\renewcommand{\theequation}{\thesection.\arabic{dummy}\alph{equation}}
\begin{eqnarray}
\addtolength{\abovedisplayskip}{\extraspaces}
\addtolength{\belowdisplayskip}{\extraspaces}
\addtolength{\abovedisplayshortskip}{\extraspace}
\addtolength{\belowdisplayshortskip}{\extraspace}}
\newcommand{\eaa}{
\end{eqnarray}
\setcounter{equation}{\value{dummy}}
\renewcommand{\theequation}{\thesection.\arabic{equation}}}

\newcommand{\ban}{\begin{eqnarray*}
\addtolength{\abovedisplayskip}{\extraspaces}
\addtolength{\belowdisplayskip}{\extraspaces}
\addtolength{\abovedisplayshortskip}{\extraspace}
\addtolength{\belowdisplayshortskip}{\extraspace}}
\newcommand{\ean}{\end{eqnarray*}}

\newcommand{\newsection}[1]{
\vspace{15mm}
\addtocounter{section}{1}
\setcounter{equation}{0}
\setcounter{subsection}{0}
\setcounter{footnote}{0}
\begin{center}
{\Large \thesection. #1}
\end{center}
\nopagebreak
\medskip
\nopagebreak}

\newcommand{\ber}{\begin{eqnarray}}
\newcommand{\eer}{\end{eqnarray}}

\newcommand{\dd}[1]{{\partial \over \partial {#1}}}

\newcommand{\LL}{\Lambda}

\newcommand{\ep}{\epsilon}
\renewcommand{\dd}{{\delta}}
\newcommand{\bb}{\beta}
\renewcommand{\aa}{\alpha}
\renewcommand{\ss}{{\sigma}}

\newcommand{\hf}{{\textstyle{1\over 2}}}
\newcommand{\tr}{{\rm tr}}

\renewcommand{\d}{{\overline{\partial}}}
\newcommand{\1}{{\it 1}}

\newcommand{\ra}{\rightarrow}

\newcommand{\lra}{{\mathop{\leftrightarrow}}}

\newcommand{\Tr}{{\rm Tr}}
\newcommand{\pr}{{\prod}}
\newcommand{\N}{{\tr \over N}}

\newcommand{\bl}{\begin{list}{({\it\roman{xxx}})}{\usecounter{xxx}}}
\newcommand{\el}{\end{list}}
\newcounter{xxx}

\begin{document}
\hsize36truepc\vsize51truepc
\hoffset=-.4truein\voffset=-0.5truein
\setlength{\baselineskip}{14pt}
\setlength{\textheight}{8.5 in}
\vphantom{0}\vskip1.1truein
\leftline{A STRING PROJECT IN MULTICOLOUR QCD}
\vskip.65truein

\hbox{\obeylines\baselineskip14pt\parskip0pt\parindent0pt\hskip1.1truein
\vbox{Vladimir Kazakov
\vskip.1truein
Laboratoire de Physique Th{\'e}orique
de l'{\'E}cole Normale Sup{\'e}rieure
24 rue Lhomond, Paris 75231, France
(permanent address:
Academy of Sciences of USSR, Moscow)
}}

\vskip .65truein
\leftline{Abstract}
{\baselineskip 12pt plus 1pt minus 1pt
\small{
\noindent

Some old and new evidence for the existence of the string (planar
random surfaces) representation of multicolour QCD are reviewed. They
concern
the random surface representation of the strong coupling expansion in
lattice multicolour gauge theory in any dimension.

Our old idea of modified strong coupling expansion in terms of
planar random surfaces, valid for the
physical weak coupling phase of the four-dimensional QCD, is explained
in detail. Some checks of the validity of this expansion are proposed.
(The lectures given in the Trieste Spring School and Workshop-1993 on
String Theory)
}}
\vskip.65truein

\newsection{Introduction}

The principal model considered in this communication will be the
Wilson lattice $U(N)$ gauge theory without quarks,
particularly its large N (multicolour) limit.

The action of the theory is defined  on the hypercubic D-dimensional
lattice with the vertices labeled by the D-dimensional vectors $x$ and
the vectors of the links as $\mu, \nu, etc.$ (so that the plaquette is
denoted as $(x,\mu,\nu)$).

\be
S[U]=N \bb \tr \sum_{x,\mu,\nu} U_{x,\mu} U_{x+\mu,\nu} U^+_{x+\nu,\mu}
U^+_{x,\nu} + \ \ \mbox{compl. conj.}
\label{lgt}
\ee

The partition function $Z$, the free energy $F$ and the Wilson
average $W(C)$ along a closed loop $C$ on the lattice
are now expressed as:

\be
Z=e^{N^2 F} = \int \pr_{x,\mu} [dU_{x,\mu}]_H \exp N \bb S[U]
\label{part}
\ee

\be
W(C) = < \N \pr_{(x,\mu)C} U_{x,\mu} >
\label{loop}
\ee
with the obvious definition of the average $<...>$.

The Haar measure $[dU]_H$ for the U(N) integrals is characterised by the
unitarity condition
\be
U^+U=U U^+ = I
\label{unit}
\ee

The famous paper of K.Wilson \cite{Wilson},
which explains the confinement of quarks in terms of
the strong coupling (SC) expansion in lattice QCD,
left a wide area in which to improve upon ;  this lattice "superconfinement"
resulted in a string tension $K(\bb)$ which was wrongly scaled to describe
the weak coupling (WC) phase.
Namely, from the Wilson calculations for the planar loop $C$
of sufficiently large size one obtains
in the (SC) limit the area law asymptotics:

\be
W(C) \sim \exp[-K(\bb) A_{min}(C) ]
\label{area}
\ee
where $A_{min}(C)$ is the minimal area of the surface spent on $C$,
with
\be
K(\bb) \sim -\log(\bb)
\label{tension}
\ee
instead of the correct physical asymptotics
\be
K(\bb) \sim \LL^2 \exp[-{48 \pi^2 \over 11} \bb]
\label{trans}
\ee
dictated by the dimensional transmutation mechanism of asymptotically
free theory.

It is widely believed, after years of computer simulations of the
lattice SU(N) gauge theory,
that in the case N=2 and N=3 gauge groups,  if we move
from small $\bb$, where (\ref{tension}) is valid, to larger $\bb$, we
meet a sharp crossover: an abrupt change in behaviour in physical
quantities as functions of $\bb$, before we reach the physical
confinement with the string tension (\ref{trans}). Nobody knows whether
we can describe at least in principle the physical phase by summing up
the SC series. Even if the answer is positive, practically, it turns out to be
impossible.

The situation becomes even worse if N  increases. The crossover
turns into a real first order phase transition, toppling any hope that
the SC expansion can be extended into the physical phase. This phase
transition corresponds to the breakdown of the Z(N) symmetry of the
centre of the gauge group. It has indeed to be broken in the physical
phase of the theory, since we know very well that each gauge matrix $U$
on the lattice has to fluctuate close to some singled out value on the
group, which leaves no space for the discrete Z(N) degrees of freedom.
This phase transition is clearly seen even in rather naive mean field
calculations.

In the limit of large N, which is our only hope for any free string
representation (if not exact solution) of QCD, a new complification
arizes, namely, the Gross-Witten (GW) phase transition \cite{gw}.
This is also
characterised by a narrow distribution of any particular lattice
gauge variable around some singled out point on the group space, once
one increases $\bb$ beyond the GW transition point. If we consider the
eigenvalues of a gauge variable (which are the gauge invariants) we can
say that their distribution in the WC phase
does not take the whole unit circle, as in
the SC phase, but is squeezed into a smaller interval. The Z(N) symmetry of
the group centre will be broken anyway.
Note that although the neighbouring gauge variables should not be far
away from each other on the group space, on the scale of the
correlation radius of the theory (in other words,
physical mass, or confinement scale), different U-variables cover the
whole group space. This property itself can be viewed as a definition
of the correlation radius.

All this looks like very bad news for the SC expansion in the Wilson
lattice theory. Even the
hope of an exact string representation of QCD would be significantly
reduced by this failure, since one of the few indications of its
existence is the representation of the SC expansion in terms of the sum
over non-interacting random surfaces on the hypercubic D-dimensional
lattice, found in \cite{Kaz1,Kaz2} and advanced in \cite{Zuber,Kos1,Kos2}
(we will present below a short discription of this construction).

Nevertheless, Wilson's original idea of confinement
on the lattice seems to be too nice to discard so easily.
A long time ago, the author proposed a possible way out of the impasse
\cite{Kaz2}.
Namely, it was suggested to modify the SC expansion by the introduction
of new weights on the links of random surfaces, which have a
non-trivial dependence on $\bb$ and must be calculated selfconsistently
from the unitarity condition (\ref{unit}) for the gauge variables. These
weights can be expressed in terms of averages
of traces of products of the Lagrange
multipliers for the constraints (\ref{unit}) on each link. The weights do
not fluctuate in the large N limit due to the factorization property,
and are uniform in the physical space in light of the translational and
rotational invariance. So, they can be considered as the effective
"world sheet coupling constants", though this notion is a bit vague on
the lattice.

The whole construction is very close to the Stanley solution of the
N-vector field theory (on the lattice) in the large N limit
(though our construction is as yet far from a solution).
Let us recall these simple and nice arguments of Stanley, to compare
with ours.

The partition function of the N-vector field $n_a(x)$ normalized as
\be
n_b^2(x)=1
\label{sphere}
\ee
is defined as follows:
\be
Z=\int \pr_x D^N n(x) \dd(n_b^2(x)-1)
\exp[ N {\bb \over 2} \sum_{x,\mu} n_b(x) n_b(x+\mu)]
\label{vector}
\ee
We can try to investigate this model in terms of the SC expansion by
expanding (\ref{vector}) in powers of $\beta$ and integrating order by
order with respect to the compact measure in $n(x)$.  The strong
coupling diagrams will look like sums over paths with the weights
$\bb^{length}$, which might have nontrivial couplings in the points of
selfintersections. Let us
demonstrate, say, the picture corresponding to the constraint (\ref{sphere}):
\be
<n_a^2(x)> =1
\label{norm}
\ee
The typical strong coupling diagram is shown in fig.1. In the large
N limit the diagrams will simply be trees of loops attached to each
other at single points in physical space.

Let us remind  ourselves how to sum up these trees.

We introduce the lagrange multiplier field $\aa(x)$ to impose the
constraint (\ref{unit}) on the integration measure, and then integrate
out the $n$-field to obtain the following functional integral:
\be
Z=\int \pr_x D \aa(x)
\exp[ N  \sum_x \aa(x) - {N \over 2}
\Tr \log\Big(-\bb (\dd_{x,x+\mu}+\dd_{x,x-\mu} + \aa(x)\Big)]
\label{mult}
\ee
In the limit of large N we can solve this problem by looking for the
translationary invariant saddle point for $\aa$,

\be
{\dd S_{eff}[\aa] \over \dd \aa(x)}=0
\label{stat}
\ee
which gives the equation:
\be
1={1 \over (2\pi)^2} \int_0^{2\pi}d^2p{1\over 2\bb \sum_\mu \cos(p_\mu) +\aa}
\label{min}
\ee

In the limit $\bb \ra \infty$ we shift the variable $\aa$
\be
\aa =  \bb (4- m^2)
\label{shift}
\ee
expand the $\cos p$ in (\ref{min}) (since we are going to get the
continuous limit) and obtain
\be
\bb={1 \over (2\pi)^2} \int d^2p{1\over p^2 +m^2} = {1 \over 4\pi}
\log{\LL^2 \over m^2}
\label{log}
\ee
which gives the well-known dimensional transmutation for the physical
mass scale:
\be
m^2 = \LL^2 e^{-4\pi \bb}
\label{transm}
\ee
where $\LL$ is the ultraviolet cutoff.

The end result is a multiplet of N scalar noninteracting particles
with the same mass
(\ref{transm}). This result is exact in the large N limit. One can easily
see that the saddle point condition (\ref{min}) corresponds to the
normalisation condition (\ref{norm}).

If we expand the formula (\ref{min}) in powers of $\bb$ we get:
\be
1=\sum_{paths P_{xx}} \big({\bb \over <\aa>}\big)^{Length(P_{xx})}
\label{path}
\ee
where $P_{xx}$ is a path starting and finishing at the same point $x$
on the lattice. Instead of summing over the whole tree of the loops we
sum over only one free closed path with the renormalised weight:
\be
\big({\bb \over <\aa>}\big)^{Length(P_{xx})}
= \big(4- m^2 \big)^{Length(P_{xx})}
\label{modif}
\ee

We see that this sum over random paths diverges at $m=0$, as it should
do, in order to have an appropriate continuous limit.

This is an example of a modified SC (MSC) expansion which works in the
physical SC phase.

It will be useful for the future comparison with QCD to show the
relationship between the lagrange multiplier field and a vector field
condensate (I thank A.Polyakov for his comments on  this subject).
We have the following equation for the correlation function
of the vector field in the background of the lagrange multiplier:
\be
\bb \sum_{\mu=\pm 1,\pm 2} <n_b(x+\mu) n_b(y)> + <\aa
n_b(x)n_b(y)>+\dd_{x,y}=0
\label{corcon}
\ee
For $x=y$ we obtain in the continuous limit:
\ba
&  <\aa>=4\bb -\bb \Lambda^{-2}\pi^2 < n_b(x)\Delta n_b(x)> - 1   \\
& = 4 \bb +  {1 \over 4 \Lambda^2} \int { d^2p p^2 \over p^2+m^2 } - 1\\
& =  4\bb -{\pi\over 4}{m^2 \over \Lambda^2} \log{\Lambda^2 \over m^2}
\label{relcon}
\ea
This is the relation between the modified weight in the sum over paths
and the vector field condensate. Of course, it agrees with (\ref{transm}).

We presented here this well-known solution of Stenley model since it
will be our guideline for the construction of the MSC
expansion in lattice QCD, which looks to be generalizable to
the physical WC phase.

\newsection{Free random surface representation of the strong coupling
expansion in D-dimensional lattice gauge theory}

In this section we are going to reformulate the standard SC expansion
of Wilson gauge theory in terms of FREE random surfaces on the D-dimensional
hypercubic lattice. This formal representation will be exact order by
order in $\bb$. The result will consist of a description of the
elementary geometrical objects (plaquettes, "saddles" et.c) from which
we construct this surface.

 To demonstrate the idea let us start from a simpler model of random
surfaces on the lattice: the Weingarten model. It has the same action
as  Wilson gauge theory, but instead of the unitary Haar measure one
takes a gaussian measure:
\be
[dU]_W = d^{2N^2}U \exp[-N \tr U^+U]
\label{wein}
\ee
It can be represented in terms of lattice random surfaces by the
standard SC expansion. We expand the exponent in
(\ref{part}) in powers of $\bb$ and perform gaussian integrals using
Wick's theorem:
\be
\int [d^{2N^2}U]_W U^+_{ij}U_{kl} = {1 \over N} \dd_{il}\dd_{jk}
\label{wick1}
\ee

\be
\int [d^{2N^2}U]_W U^+_{i_1j_1}U_{k_1l_1} U^+{i_2j_2}U_{k_2l_2}
= {1 \over N^2} \dd_{i_1l_1}\dd_{j_1k_1} \dd_{i_2l_2}\dd_{j_2k_2}
+ {1 \over N^2} \dd_{i_1l_2}\dd_{j_1k_2} \dd_{i_2l_1}\dd_{j_2k_1}
\label{wick2}
\ee
and so on.

Geometrically this means that we can glue plaquettes together by means of
these integrations over the common link variables, as is shown in fig.2
for (\ref{wick1}) for two plaquettes, and in fig.3 for (\ref{wick2})
for four plaquettes. Continuing this process we arrive at closed planar
surfaces built on the hypercubic lattice. Due to the matrix structure
of the theory, every connected piece of the surface will be weighed
with the weight $N^{2-2G}$, where $G$ is its genus. this factor is
explained
in the standard way, for matrix models. In the Weingarten model
the 'tHooft limit of large N leads precisely to planar surfaces.
Due to the equal weights of different terms in
the formulae (\ref{wick1})-(\ref{wick2}), there are no extra weights  at
the intersections of surfaces which means the surfaces are the world
sheets of a ``free string''.

Finally, the SC expansion for  the free energy in
the Weingarten model can be represented
in the large N  limit as a sum over planar surfaces $\ss_W$ on the
hypercubic lattice:
\be
F(\bb)= \sum_{\ss_W} \bb^{A(\ss_W)}
\label{cub}
\ee
where $A(\ss_W)$ is the area of the surface (the number of plaquettes
from which it is built).

It is known  that for any dimension higher than 1 this model has
a pathological behaviour and degenerates into tree like
configurations at the critical point \cite{frohlich}.

What will change if we take instead of  (\ref{wein}) the U(N) Haar
measure of the Wilson gauge theory? Nothing will be different for the
gluing of two plaquettes, since the integral (\ref{wick1}) will be the
same. However, for four plaquettes, we will have instead of
(\ref{wick2}):
\ba
& \int [d^{2N^2}]_H U^+_{i_1j_1}U_{k_1l_1} U^+{i_2j_2}U_{k_2l_2}
= {1 \over N^2-1} [\dd_{i_1l_1}\dd_{j_1k_1} \dd_{i_2l_2}\dd_{j_2k_2}
 + \dd_{i_1l_2}\dd_{j_1k_2} \dd_{i_2l_1}\dd_{j_2k_1}]  \\
& - {1 \over N(N^2-1)} [\dd_{i_1l_1}\dd_{k_1j_2} \dd_{i_2l_2}\dd_{k_2j_1}
-  \dd_{i_1l_2}\dd_{k_2j_2} \dd_{i_2l_1}\dd_{k_1j_1}]   \\
& = {1 \over N^2} \dd_{i_1l_1}\dd_{j_1k_1} \dd_{i_2l_2}\dd_{j_2k_2}
- {1 \over N^3} \dd_{i_1l_1}\dd_{k_1j_2} \dd_{i_2l_2}\dd_{k_2j_1}   \\
& + {1 \over N^4} \dd_{i_1l_1}\dd_{j_1k_1} \dd_{i_2l_2}\dd_{j_2k_2}
- {1 \over N^5} \dd_{i_1l_1}\dd_{k_1j_2} \dd_{i_2l_2}\dd_{k_2j_1}+...
+ (i_1,j_1 \lra i_2,j_2)
\label{saddle}
\ea
The first  line in the r.h.s. of (\ref{saddle}) represents
the original Wick contractions of (\ref{wick2}), whereas the second
corresponds to new, cyclic contractions of indices (and
plaquettes), etc.

In the large N limit, we have to expand the N-dependent coefficients in
(\ref{saddle}) in powers of 1/N and interprete each term geometrically in terms
of pieces of a free planar random surface on the D-dimensional
hypercubic lattice. The most natural interpretation is presented
graphically in fig.4. The first term represents one of Wick contractions of
indices (the rest of the gauge variables on these four plaquettes is
contracted into two separate traces at the boundaries of two
disconnected couples of plaquettes).

The second term corresponds to the
cyclic coupling of indices (the rest of the gauge variables are
contracted in a single trace around a single boundary). This term looks
like a saddle (though quite a singular one, with zero radius of
curvature). Topologically it is a disc which is glued into the
random surface. It has the an extra 1/N power with respect to the
previous term, corresponding to smaller Euler characteristics $\kappa$
(it is equal to 1 for each disc). In this way the topological expansion of
'tHooft attaches to our surfaces the factor:
\be
N^{2-2G}=N^\kappa
\label{topol}
\ee

The saddles seem to play an important role in the
whole construction. Note that the sign is negative in front of this
term.

The last term in fig.3 represent the next term of the 1/N expansion of the
coefficient in the first term in the l.h.s. of (\ref{saddle}) and can
be described as a tube connecting two couples of plaquettes. It
again has the topology of a disc.

All these terms can in principle contribute in the large N limit.
Higher order
terms in the 1/N expansion in (\ref{saddle}) describe higher order
topologies.

This procedure can be continued for higher n-correlators of gauge
matrices of the type
\be
\int [d^{2N^2}]_H U^+_{i_1j_1}U_{k_1l_1}..... U^+{i_nj_n}U_{k_nl_n}
\label{corr}
\ee
These correlators, being calculated in the same manner, give all
possible connected or disconnected objects like multiple "saddles",
drawn in fig.5, and give rise to a cyclic contraction of indices in
(\ref{corr}), like
\be
\dd_{i_1l_1}\dd_{k_1j_2} \dd_{i_2l_3}....\dd_{k_nj_1}
\label{nsaddle}
\ee
as well as tubes and their mixtures, which can be build from 2n
plaquettes.
Every connected part of such
a configuration is accompanied by some numerical coefficient which
corresponds to appropriate index contractions in the integral
(\ref{corr}).

This interpretation of the strong coupling expansion
in terms of random lattice surfaces was proposed in
\cite{Kaz1,Kaz2}. The factors corresponding to the n-saddles,
consisting from cycling gluings of 2n plaquettes, were found there to be
the Catalan numbers $f_n$ (see next section for their calculation):
\be
f_n=-(-1)^n {(2n)! \over 2(2n-1) (n!)^2 }
\label{cata}
\ee
so that
\be
f_1=1, \ \ \ f_2=-1, \ \ \ , f_3=2, \ \ \ f_4=-5, \ \ \ f_5=14
\label{ex}
\ee

The whole variety of coefficients corresponding to this zoo of objects
was calculated in \cite{Zuber}. Let us comment also that the rules for
the sum over surfaces found here for the free energy, are directly
generalisable to the Wilson average W(C). One can consider the gauge
variables forming the loop factor in (\ref{loop}) as the edge of a
surface, to which the plaquettes can be attached either by Weingarten
type contractions, or by saddles, tubes, et.c. The Wilson average can
be viewed as an open string amplitude.

Fortunately, as was shown by Kostov
\cite{Kos1,Kos2}, with the exception of the multiple saddles, these
complicated objects can be ignored and the sum over surfaces can be
reduced to a sum over one-link-irreducible surfaces with multiple
saddles. It was shown by Kostov that any surfaces which can be cut
into two pieces by cutting along a single link (one-link-reducible
surfaces) cancel each other due to sign changing factors.

This theorem of Kostov is a direct consequence of the
XSunitarity of the gauge variables (\ref{unit}). Let us sketch the proof
of it (the details can be found in (\cite{Kos2}).

One can use
the "backtracking" condition for an arbitrary Wilson average
(\ref{loop}): if we cut the contour C in a point and glue in this cut a
path consisting from two links $l*l^{-1}$ going there and back in the same
direction $\mu$ on the lattice, the Wilson average will not change (see
fig.5):
\be
W(C*l*l^{-1}) = W(C)
\label{backtr}
\ee
Now consider dressed saddles: by definition these correspond to the sum
of all surfaces attached to the link on which the  saddle is
situated. Namely, it is the bare saddle, considered above, plus all
surfaces connected to it by "tubes" (also considered above). We can
have in the sum over surfaces in the l.h.s. of (\ref{backtr}) three
 situations:
a dressed link vertex of the n-th order can be attached to both links
$l,l^{-1}$,
or two link vertices of the orders (n-k) and k can be attached
separately to each link, or there will be no plaquettes attached to
these links at all (which corresponds precisely to W(C) without back
tracking). So, introducing the weights $F_k$ of dressed saddles, we
obtain from (\ref{backtr}):
\be
\sum_n W_n(C)[F_n + \sum_{k=0}^n F_{n-k} F_k ] + W(C) = W(C)
\label{proof}
\ee
where $W_n(C)$ is the sum over surfaces spanned on the contour C with
dressed n-saddle attached to two backtracking links.

Since C can be arbitrary here, we conclude that
\be
F_n + \sum_{k=0}^n F_{n-k} F_k  = 0
\label{catdef}
\ee
which defines the catalan numbers (\ref{cata}). From here we
conclude that the weights of the dressed saddles are also equal to the
Catalan numbers: $F_n=f_n$. Hence, we can throw away the one-link-
irreducible surface, together with the objects more complicated then
multiple saddles, from the sum over surfaces.

As usual, the partition function corresponds to the sum of various
disconnected random surfaces, each having the topology of a sphere. Since
these surfaces are non-selfinteracting (there is no excluded volume
problem for them) we can be sure that the free energy corresponds to
the sum over only connected random surfaces.

Finally, the random surface representation of the SC expansion in the
Wilson gauge theory looks quite simple:
\be
F(\bb)=\sum_{\ss} \bb^{Area(\ss)} \prod_{s \ss} f_{n_s}
\label{surf}
\ee
where $\ss$ are the one-link-irreducible planar surfaces built from
plaquettes glued together in saddles denoted by $s$,
 on the D-dimensional hypercubic lattice. $f_{n_s}$ are the Catalan
factors attached to these saddles (we consider normal contraction of
two plaquettes as a saddle of order 

The analogous representation for $W(C)$ can be given in terms of a sum over
surfaces spanned on a loop $C$.
Very few things are known about the critical behaviour of this SC
expansion in multicolour Wilson gauge theory. The no-go theorem of
\cite{frohlich} is not directly applicable here, since we have
sign changing terms. It might or might not have a nontrivial critical
behaviour, but its direct continuation to the physical WC phase seems
quite unprobable in virtue of one of the two phase transitions mentioned
in the introduction.

Can we modify the random surface representation (\ref{surf})
in order to describe the WC phase of QCD as well? We will propose a
possible modification in the next section, and then we will give some
arguments in favour of it.

\newsection{Modified strong coupling expansion for the physical phase
in $QCD_4$}

To explain the idea of the modified SC expansion, proposed in
\cite{Kaz2} and tested on some examples in \cite{Tol}, we consider two
similar, but technically slightly different approaches.

The first one, originally proposed in \cite{Kaz2}, leads to a simpler
geometrical picture for random surfaces, whereas the second based on
the representation introduced in \cite{Kos2} leads to a more promising
quantitative scheme.

\subsection{First construction for modified SC expansion}
In the first case we introduce for each link gauge variable $U_\mu$ a
hermitean lagrange multiplier matrix $\aa_\mu$, parametrizing the Haar
measure in one of two following ways:
\ba
& [dU]_H=d^{2N^2}U \dd^{N^2}(U^+U-I) =   \\
&\int d^{N^2}\aa e^{N \tr \aa(I-U^+U)}=   \\
&\int d^{N^2}\aa e^{N \tr \aa(I-UU^+)}
\label{param}
\ea
It is our choice whether to take the second or the third line as a
definition of $\aa$ for any link variable. Let us choose it in the most
symmetric way: we classify all the vertices of the D dimensional
lattice in chess order, as even and odd, and we take the
definition of $\aa_\mu$ according to the second line of (\ref{param}), if
the corresponding link goes from an even to an odd vertex (in the
positive direction of the coordinate axes $\mu$), and the line three of
(\ref{param}), if otherwise.

Now we have a double functional integral for (\ref{part}): in
hermitean matrices $\aa$
and in complex matrices  $U$. Let us perform first the integral over
$U$'s by means of strong coupling expansion in $\bb$. In each order the
integrals are now purely gaussian, with the matrix propagators equal
to
\be
\int d^{2N^2} e^{N\tr \aa U^+U} U^+_{ij}U_{kl} =
{1 \over N} (\aa^{-1})_{il}\dd_{jk}
\label{prop1}
\ee
for the "even" links, and
\be
\int d^{2N^2} e^{N\tr \aa UU^+} U^+_{ij}U_{kl} =
{1 \over N} \dd_{il}(\aa^{-1})_{kj}
\label{prop2}
\ee
for the "odd"links.

The integration over $U$'s can be performed according to the same rules
as for the standard Weingarten model
with the measure (\ref{wein}). We will get the sum over exacly
the same hypercubic surfaces $\ss_W$ as in Waingarten model, but since
the propagators are now modified, new $\aa$-depending weights should
be attached to every even vertex $es$ of the surface:
\be
F(\bb)=\sum_{\ss_W} \bb^{Area(\ss_W)}
\prod_{es_W \ss} g_{\{\mu\}_{es}}(\bb)
\label{wsurf}
\ee
where
\be
g_{\{\mu\}}=
g_{\mu_1\mu_2...\mu_n}(\bb)=
<{1 \over N}\tr[\aa_{\mu_1} \aa_{\mu_2} ... \aa_{\mu_n}]^{-1}>
\label{gfact}
\ee
where $\mu_1, \mu_2,...,\mu_n$ are the directions, either
positiv or negativ, of links around a given even vertex on the surface).

Instead of futher integration over $\aa$'s,
we have already substituted in (\ref{gfact})
the emerging traces of products of
$\aa_\mu^{-1}$'s by their averages, according to the large N
factorization theorem \cite{Mig,Wit}. We see from this construction,
that we can formally represent any physical quantity in multicolour
lattice QCD as a sum over random hypercubic surfaces
(Weingarten type surfaces in this case) with special factors attached
to the vertices of surfaces. In virtue of  translational and
rotational invariances of any physical averages, the $g$-factors should not
depend on the position of the vertex of the surface in the
D-dimensional lattice, but only on the sequence of the links
$\mu_1,...,\mu_n$ around a given vertex of the surface (up to the
obvious rotations). So
$g_{\{\mu\}}$ play the role of "string coupling constants", though this
notion is quite vague for the lattice strings.

These constants are in fact non-trivial functions of $\bb$ and should
be calculated separately. One can propose the following formal method
for it.

In the large N limit, we can in principle define all g-factors from the
chain of obvious equations, following from the unitarity of gauge
matrices:
\be
< {1 \over N} \tr  U_{\mu_1}^+U_{\mu_1}...U_{\mu_n}^+U_{\mu_n} > = 1
\label{uncond}
\ee
where $\mu_1,...,\mu_n$ is any sequence of directions of links around
an even vertex on the lattice. The l.h.s. of (\ref{uncond}) is just a
wilson average for a loop $C\{\mu\}$ consisting of only backtrackings along the
links surrounding a given vertex. If we re-express these Wilson averages
in terms of the same sums over random surfaces spanned on these
contours, as it was done for the free energy (\ref{wsurf}), we
obtain the following conditions on $g$'s:
\be
\sum_{\ss_W, \partial\ss_W = C\{\mu\}} \bb^{Area(\ss_W)}
\prod_{es \ss_W} g_{\{\nu\}_{es}}(\bb) =1
\label{surcond}
\ee
where the sum is taken over all the surfaces having as a boundary the
abovemantioned backtracking contour $C\{\mu\}$.

This chain of nonlinear equations can serve at least
in principle for the calculation of the string couplings $g_{\{\nu\}}(\bb)$.
Note the similarity (not accidental) with the equation
(\ref{path}) for the N-vector field: there the sum over paths pinned to
a point in the physical space, with the renormalized hopping parameter
$\bb/\aa$ was equal to 1, as a consequence of the normalization of the
vector field, where as here the sum over random surfaces pinned to a
sequence of links around a vertex on the lattice, with renormalized
string couplings, is equal to 1, as the consequence of unitarity.

Of course, technically all this looks too complicated. There are too
many factors $g$ to calculate. Therefore we use this approach only to
demonstrate that the sum over random surfaces is quite a natural
representation of the Wilson multicolour QCD.

Why should this modified strong coupling expansion work in pnysical WC
phase of the theory?
Our hope in this approach is that the equations (\ref{surcond}) can
have two different branches of solution: the SC branch, which
corresponds to the standard $\bb$ expansion considered in the previous
section and valid up to some critical $\bb_c$, and the physical WC
branch valid beyond $\bb_c$. It is obvious that in the SC phase the
g-factors will be just $\bb$-independent numbers, where as in the WC
phase they should be nontrivial functions of $\bb$. The traces of
$\aa$-matrices serve here as an order parameter for the corresponding
phase transition (of Gross-Witten or $Z_N$ breaking type).
We will show in the next section that this modified SC expansion works
indeed in both phases in the simplest example: the one plaquette model.

This construction is hardly useful for practical calculations, but the
existence of a simple random surface representation of Weingarten type
suggests that the search for the continuous QCD-string is not in vain.

\subsection{Second construction for modified SC expansion}

Let us describe the second construction for the modified SC expansion.
The corresponding surfaces in this construction
will be more complicated, including now the n-saddles. However
technically it will be much more tractable since we will effectively obtain a
sequence of weights $g_n{\bb}$ (traces of only one $\aa$-matrix),
attached to n-saddles
labeled by only one integer $n$.

According to the trick proposed in (\cite{Kos2}), let us double every
gauge variable on each link $l$:
\be
U_l \ra U_lV^+_l
\label{double}
\ee
We can integrate now in each of the matrices separately. This is the
same Wilson gauge theory, since the matrix $V$
can be easily absorbed into $U$, in virtue of the invariance of the
Haar measure.

Now the Haar measure on every link can be parametrized as:
\be
 [dU]_H [dV]_H =   \\
\int d^{N^2}\aa_1
\int d^{N^2}\aa_2
\int d^{2N^2}U
\int d^{2N^2}V
e^{N \tr \aa_1(I-U^+U)}
e^{N \tr \aa_2(I-V^+V)}
\label{dparam}
\ee
We are ready now to integrate out the $U,V$ variables by means of the
formal $\bb$-expansion with fixed $\aa$'s. Since each of the link
variables is doubled we have to apply the Wick theorem separately to
U-half-link and V-half-link independently. Let us consider two
examples of one link integrals over $U$ and $V$ with $\aa$ fixed:
with two link-variables:
\be
<(UV^+)_{ij}(VU^+)_{kl}>^{(0)}_{U,V}
= {1 \over N} \dd_{il}\dd_{jk} \N \aa^{-2}
\label{2plaq}
\ee
and with four link-variables:
\ba
&  <(VU^+)_{i_1j_1}(UV^+)_{k_1l_1}
(VU^+){i_2j_2}(UV^+)_{k_2l_2}>^{(0)}_{U,V}  \\
&= {1 \over N^2}(\N \aa^{-2})^2 \dd_{i_1l_1}\dd_{j_1k_1}
\dd_{i_2l_2}\dd_{j_2k_2}\\
& +{1 \over N^3}\N \aa^{-4}  \dd_{i_1l_2}\dd_{k_2j_2}
\dd_{i_2l_1}\dd_{k_1j_1}\\
& + (i_1,j_1 \lra i_2,j_2)
\label{4plaq}
\ea
where we introduced the hermitean matrix
\be
\aa^2=\aa_1^\hf \aa_2 \aa_1^\hf
\label{alpha}
\ee
So the cyclic contraction of indices corresponding to a saddle,
appears here in a natural manner, even on the phase of gaussian
integrations. To every n-saddle a factor
\be
g_n(\bb)=< \N \aa^{-2n}>
\label{sfact}
\ee
should be attached. As in the previous construction, in virtue of
translational and rotational invariance, and the large N factorization
theorem, we can already take the average for each factor $g_n$,
which will not depend on the position or orientation of the link. So
$g_n$'s are   nontrivial functions of $\bb$, labeled by only one
integer $n$. The corresponding sum over surfaces $\ss*$ will consist
from surfaces built from plaquettes glued together by saddle-like
configurations $s$:
\be
F(\bb)=\sum_{\ss*} \bb^{Area(\ss*)} \prod_{s } g_{n_s}
\label{msurf}
\ee
Note that unlike the surfaces $\ss$ in (\ref{surf}) the surfaces
$\ss^*$ emerging here can be one link reducible.

The random surface picture here is more complicated then in the
previous construction, and a good question is whether we can describe
the emerging saddles as special vertex operators in a continuous string
theory. We will discuss this possibility in the concluding section.

Now we have to calculate the weights $g_n$. The best way to do it is to
write down the effective action for two matrix variables $\aa_1,\aa_2$
(as in the previous construction, they will not depend either on the space
coordinates, or on orientations in the large  N limit)
defining these factors through eq. (\ref{sfact}):
\be
S_{eff}(\aa_1,\aa_2) = N \tr [\aa_1+\aa_2-\log[\aa_1\aa_2]] + N^2 F[\aa^1\aa_2]
\label{seff}
\ee
where $F[\aa_1\aa_2]$ is the same as $F$ in (\ref{msurf}), but with
unaveraged traces of $\aa$'s instead of $g_n$'s.

The partition function of QCD can be written as a two matrix problem:
\be
Z(\bb)=\int d^{N^2}\aa_1\int d^{N^2}\aa_2 e^{S_{eff}(\aa_1,\aa_2)}
\label{zz}
\ee

We propose here to reduce the double integral in $\aa_1,\aa_2$ to an
integral over the only matrix variable (\ref{alpha}). For this purpose
one has to integrate out the extra degrees of freedom.
Introducing the variable (\ref{alpha}) and
\be
\gamma=\aa_1
\label{gamma}
\ee
we rewrite (\ref{zz}) as
\be
Z(\bb)=\int d^{N^2}\aa^2\int d^{N^2}\gamma \exp \Big(N\tr
[\gamma+\gamma^{-1}\aa^2 - \log\gamma] + N^2 F[\aa^2] \Big)
\label{zzz}
\ee
Let us compare it with the well-known  integral  over
the unitary matrix $U$ in an external field $\aa^\hf$:
\be
e^{N^2 B[\aa]} =  \int [dU]_H e^{N\tr\aa^\hf [U + U^+]}
\label{BG}
\ee
By introducing the parametrization
\be
 [dU]_H=d^{2N^2}U \dd^{N^2}(U^+U-I) =
\int d^{N^2}\gamma e^{N \tr \gamma (I-U^+U)}
\label{gparam}
\ee
and integrating out $U$ we can see that the integral over $\gamma$ in
(\ref{zzz}) is precisely given by (\ref{BG}):
\be
Z(\bb)=\int d^{N^2}\aa^2 \exp N^2\Big(
 B[\aa] + F[\aa^2] \Big)
\label{zzzz}
\ee

One can find in \cite{BrG} the expression for both the SC and WC phases of
this integral. We need only the WC branch of this solution, found in
\cite{Nauen}:
\be
B[\aa] = 2/N \sum_{k=1}^{N} \aa_k - {1 \over 2N^2}\sum_{k>j} log[\aa_k+\aa_j]
-{3\over 4}
\label{wcbg}
\ee
where $\aa_1,...\aa_N$ are the eigenvalues of the $\aa$-matrix.

One can easily obtain this result by using the representation
$U=e^{iA}=1+iA-\hf A^2+...$ of the unitary matrix in (\ref{BG}) and
keeping only guassian terms. This approximate calculation gives
nevertheless the exact result in the WC phase, as is the case for the
matrix integral in \cite{IZ}.

Finally, we can define the momenta (\ref{gfact}) as the saddle point
condition on the density of eigenvalues $\rho(\aa)$ in the one-matrix
integral (\ref{zzzz}):
\be
-1+ \sum_{n=1}^{\infty} \aa^{-2n-1} F_n[g_1,g_2,...] =
P\int d\mu \rho(\mu) \Big( {1 \over \aa-\mu} + \hf {1 \over
\aa+\mu}\Big)
\label{spe}
\ee
where $F_n$ are the sums over the surfaces built according the same
rules as (\ref{msurf}), but attached to a fixed saddle of the order n.
So to compute $g_n$'s we have first to calculate these sums over surfaces
(the most nontrivial part of the problem) then find $\rho(\aa)$ from
(\ref{spe}) as a function of $g_n$'s and then to find them from the
selfconsistency condition:
\be
\int d\aa \rho(\aa) \aa^{-2n} = g_n(\bb)
\label{selfc}
\ee

One can recognize the geometrical similarity of the eq. (\ref{spe})
with the eqs. (\ref{path}), (\ref{surcond}).

The idea of the modified SC here is similar to the previous
construction: we hope that the factors $g_n(\bb)$ will have two
different branches as functions of $\bb$. In the SC phase they are just
$\bb$-independent numbers, where as in the physical WC phase their behaviour
changes and they turn out to be nontrivial functions of
$\bb$. We will find these two branches for the simple one-plaquette
model in the next section, and then we will demonstrate their existence
in the 4D Wilson theory.

\newsection{Simple examples: one-plaquette model}

In this section we will check the idea of the modified SC expansion on
a simple example: one-plaquette model, which was defined and solved in
\cite{gw}.
Its partition function is:
\be
Z_P(\bb) =  \int [dU]_H e^{N\tr\bb [U + U^+]}
\label{GW}
\ee
We  will use a method for its solution  explaned in \cite{Tol}
which recalls our first construction. Namely we parametrize the Haar
measure in (\ref{GW}) by the lagrange multiplier $\aa$, as in
 (\ref{param}), and after the integration over the complex matrix $U$
we obtain the following 1-matrix model:
\be
Z_P(\bb) =  \int d^{N^2}\aa e^{N\tr[\aa-\log\aa + \bb^2 \aa^{-1}]}
\label{opla}
\ee
By standard methods \cite{BIPZ} we obtain the integral equation for the
density of eigenvalues $\rho(\aa)$:
\be
-1+{1 \over \aa} + \bb^2 \aa^{-2}  =
P\int d\mu \rho(\mu)  {2 \over \aa-\mu}
\label{inteq}
\ee
If we introduce the analytic (outside the cut) function
\be
F(\aa)= P\int d\mu \rho(\mu)  {2 \over \aa-\mu}
\label{anf}
\ee
which satisfies the condition:
\be
F(\aa) \ra_{\aa \ra \infty}   2/\aa
\label{as}
\ee
and has no singularities at $\aa=0$, with its imaginary part being
equal to $\pi \rho(\aa)$, we obtain two solutions for it, separated by
the Gross-Witten phase transition at $\bb_c=1/2$. The SC solution is
valid only for $\bb < 1/2$ and corresponds to the density
\be
\rho(\aa)={1 \over 2 \pi i} (1 + {1 \over \aa})
\label{scr}
\ee
The cut is collapsed to a point here, and the calculation of momenta of
$\aa$ should be understood as a contour integral around the origin. So
we obtain:
\ba
<\N \aa^{-n} > = \oint d\aa \rho(\aa) \aa^{-n} &= 1 , \ \ \mbox{if } n=0,1 \\
                                               &= 0 , \ \ \mbox{for other } n
\label{scres}
\ea

It easy to obtain the following expression for the plaquette averages:
\be
W_n(\bb) = <\N (U^n + U^{+n})> = 2\bb^n <\N \aa^{-n}>
\label{plaqav}
\ee

We conclude from the two last formulae that:
\be
W_1(\bb) = \bb
\label{ww}
\ee
and all other $W_n$ are equal to zero in the SC phase, as it should be.

In the WC phase ($\bb > 1/2$) we find the solution for $F(\aa)$ in the
form
\be
F(\aa) = -1 +  {1 \over \aa} +{\bb^2 \over  \aa^2}+
({1 \over \aa}+{\bb \over \aa^2}) \sqrt{-(\aa^2 + 2(1-
\bb)\aa+\bb^2)}
\label{ff}
\ee
Comparing it with  the expansion
\be
F(\aa) = \sum_{k=1}^\infty \aa^k <\N \aa^{-k-1}>
\label{expa}
\ee
we can obtain:
\ba
&W_1 = 1- {1\over 4\bb}     \\
&W_2 = (1- {1\over 2\bb})^2, \mbox{etc.}
\label{www}
\ea
We see that unlike the SC phase, all $W_n$ are functions of $\bb$
tending to 1 as $\bb \ra \infty$, as it should be.

It is easy to see that in the SC phase the large $\bb$ asymptotics of
the momenta of $\aa$ are always:
\be
<\N \aa^{-n} >  \ra_{\bb \ra \infty} \bb^{-n}
\label{asmom}
\ee
which is similar to the asymptotics of a single lagrange multiplier of
the vector field, as one sees from (\ref{shift}).

It is curious to see how the unitarity relations $(U^+U)^n=I$ are satisfied
here. Say, from the integral over $U$ with fixed $\aa$ we have
\be
<\N (U^+U)> =  <\N\aa^{-1}> + \bb^2 <\N\aa^{-2}> =1
\label{unita}
\ee
One can observe that the WC solution and the SC solution
satisfy (\ref{unita}) in very different ways.

Of course the example considered here is the simplest possible one: the
modified SC expansion consists here only from one term: in (\ref{opla})
this is the last term in the exponent, where the SC expansion
parameter $\bb^2$ is modified by the weight $<\N \aa^{-1}>$. In the SC
phase this weight is just one, and $\bb^2$ corresponds to the area of
the "surface" consisting from two plaquettes. In the WC phase the
weight is already nontrivial, as you see from (\ref{www}). The same is
true for the "Wilson loops" $W_n(\bb)$.

One might think that the situation considered in this example is too
simple to test the validity of the modification of SC, since we have
only one surface in the sum over lattice surfaces here.
Let us try another formulation of this model which already has an  infinite
sum over the "surfaces" (built on one single plaqutte from an
arbitrary number of copies of this plaquette). Namely let us test our
second construction of the previous section. The doubling of the
matrice $U$ gives instead of (\ref{GW}) the two unitary matrix
integrals:
\be
Z_P(\bb) =  \int [dU]_H \int [dV]_H e^{N\tr\bb [UV^+ +VU^+]}
\label{GW2}
\ee
Introducing the lagrange multipliers as in (\ref{dparam})
and integrating over $U,V$ we obtain the sum over surfaces with
modified weights in  (\ref{zzzz}) in the form:
\be
F(\aa)= - \N \log(\aa^2-\bb^2) =
-\log \aa^2 + \sum_{k=1}^\infty {\bb^2k \over k} \N
\aa^{-2k}
\label{sos}
\ee

The saddle point equation (\ref{spe}) now reads:
\be
-1+{\aa \over \aa^2 - \bb^2}   =
P\int d\mu \rho(\mu) \Big( {1 \over \aa-\mu} + \hf {1 \over
\aa+\mu}\Big)
\label{spequ}
\ee

This equation seems to be rather nontrivial, but nevertheless it can be
easily solved in the relevant WC phase. Let us shift the matrix
variable $\aa$, defining a new variable $\ep$:
\be
\aa=\bb+\ep
\label{new}
\ee
The eq.(\ref{spequ}) now reads:
\be
-1+{\ep+\bb \over 2\ep \bb+ \ep^2}   =
P\int d\ep' \rho(\ep') \Big( {1 \over \ep-\ep'} + \hf {1 \over
2\bb+\ep+\ep'}\Big)
\label{shieq}
\ee

We can try now to solve this equation by expanding the second term in
the l.h.s. and the second term in the r.h.s. in powers of $\bb^{-1}$.
In the r.h.s. we will obtain in this way the momenta $<\ep>$ as the
coefficients of expansion. It happensto be selfconsistent to set all
of them to zero.
\be
<\ep^n>=0, \ \ \ \mbox{for} \ \ n=1,2,3,...
\label{zero}
\ee
Then the equation (\ref{shieq}) turns out to be
\be
-1+{1 \over 2\ep }   =
P\int d\ep' \rho(\ep')  {1 \over \ep-\ep'}
\label{triv}
\ee
with the solution:
\be
\rho(\ep)={1 \over 2 \pi i} (1 + {1 \over 2 \aa})
\label{wcso}
\ee
As in the eq.(\ref{scr}) the corresponding integrals should be
understood as contour integrals around the origin.

It seems to be strange that the  solution (\ref{wcso}) does not depend
on $\bb$ and is very similar to the SC one (ref{scr}). Nevertheless it
reproduces the correct WC behaviour of Wilson averages, say,
\be
W_1(\bb) = <\N (UV^+ + VU^+)> =<\N {\bb \over \aa^2-\bb^2}>=
\oint d\ep \rho'(\ep) { 1 \over 2\ep + \ep^2} = 1-{1 \over 4\bb}
\label{oppa}
\ee

We see from the last formula, that the modified SC expansion contains
here an infinite number of terms:
\be
W_1(\bb)  =\bb \sum_{k=0}^{\infty} \bb^2k <\N \aa^{-2k-2}>
\label{msce}
\ee
At $\bb=\infty$ this series diverge since $\aa \sim \bb$. This
divergence is situated at $\ep=0$, therefore it gives a necessary
contribution to the integral in (\ref{oppa}). For every particular term
in the expansion (\ref{msce}) we would loose this singularity,
therefore to get the correct weights we have to regularise the
singularity in the density of eigenvalues (\ref{wcso}).

One of the regularizations would be  the above  mentioned expansion of
(\ref{shieq}) in $1/\bb$ up to a finite number M of terms and keeping the
corresponding momenta (\ref{zero}) nonzero for a while. Then we can solve
the resulting equation by standard methods and define the inverse
momenta $<\N \aa^{-k}>$ (the weights of the "surface")
from the self-consistency condition. The "sum over surfaces" with these
weights will reproduce the correct result (\ref{oppa}) for sufficiently
big M. This method was checked numerically
in \cite{Tol} on a similar construction
and the convergency with M turned out to be very fast.

Anyway, this example shows that the modified SC expansion works well in
the WC phase of the
one-plaquette model and obeys the following general features which
should hold in more realistic models:

1.There exist two different branches (SC and WC) of the solution for the
weights of the modified SC expansion: they are numbers in the SC
phase, and non-trivial functions of $\bb$ in the WC phase.

2.The matrix lagrange multiplier has the asymptotic behaviour:
\be
\aa \sim  \bb,\ \ \mbox{for}\ \  \bb \ra \infty
\label{aas}
\ee
tending to cancel the big factor $\bb^{Area}$ in the sum over surfaces.

3.The corresponding sum over surfaces diverges at the point
$\bb=\infty$ and needs to be regularized.

\newsection{The existence for the modified weights of random surface in
physical phase of $QCD_4$}

The problem of calculating the sum over
surfaces in a physical four dimensional situation in the formulae
(\ref{msurf}) is non-trivial. Even its reformulation in terms of a
continuous functional integral of some string theory with a definite
two dimensional world sheet action is far from being achieved.

Our task here is rather to demonstrate the possibility of the existence of
such a representation in principle, then to get a well elaborated
quantitative approach to the multicolour QCD.

Namely, we will show the relationship between gluon condensates and the
momenta of the lagrange multiplier matrix.

Let us vary the action (\ref{lgt}) with the Haar measure parametrized
as in (\ref{dparam})
with respect to the fields $U^+\dd U^+, V\dd V$. We
get the following matrix operatorial equations of motion:
\be
\aa_{1 x,\mu}= \bb  \sum_\nu (UV^+)_{x,\nu} (UV^+)_{x+\nu,\mu}
(VU^+)_{x+\mu,\nu}
(VU^+)_{x,\mu}
\label{Uvar}
\ee
and
\be
\aa_{2 x,\mu}= \bb  \sum_\nu (UV^+)_{x,\mu} (UV^+)_{x+\mu,\nu}
(VU^+)_{x+\nu,\mu}
(VU^+)_{x,\nu}     \\
\label{Vvar}
\ee
Since we know that any physical gauge invariant quantity depends only
on the combination $\aa_1\aa_2$ of these matrices, we find from last
equations:
\ba
& \aa=\aa_1 \aa_2 =   \\
&   \bb  \sum_{\nu_1,\nu_2} (UV^+)_{x,\nu_1} (UV^+)_{x+\nu_1,\mu}
(VU^+)_{x+\mu,\nu_1}
(UV^+)_{x+\mu,\nu_2} (VU^+)_{x+\nu_2,\mu}
(VU^+)_{x,\nu_2}  \\
\label{chear}
\ea

The product of matrices in the r.h.s. runs along the 6 link boundary of
two plaquettes attached to each other along a fixed link $\mu$. This
"chair-like" configuration of plaquettes was used extensively for
various improvements of the Wilson action in lattice computer
simulations of QCD.

So  any momentum of the matrix $\aa$ which acquires the space and
direction independent average in the large N limit, can be represented
as an average of the power of the combination in the r.h.s. of
(\ref{chear}). Any positive power looks like a saddle configuration
of the n-th order (built from 2n plaquettes) presented in fig.4:
\ba
  <\N \aa^n> = &\bb^{2n} <\N \Big(\sum{\nu_1,\nu_2}
 U_{x,\nu_1} U_{x+\nu_1,\mu}
U^+_{x+\mu,\nu_1}
U_{x+\mu,\nu_2} U^+_{x+\nu_2,\mu}
U^+_{x,\nu_2}\Big)^n>   \\
&  + \ \ \mbox {contact terms}
\label{chese}
\ea

If we represented the n'th power in
 the r.h.s. as a sum of individual terms it
would be expressed through Wilson loops running
around the boundaries of  saddles of the n-th order (we restored the
standard Wilson gauge variables $U$, without the doubling
on each link).

For negative $n$'s (\ref{chese}) gives a local
expression of the weights of the modified SC expansion
in  (\ref{msurf}). Again, the situation is very similar here to the
calculation of the lagrange multiplier through the local condensate of
the vector field, give in the Introduction.

Let us discuss a possible continuous limit for the expression
(\ref{chese}). it is clear that for $\bb \ra \infty$ any wilson loop
tends to 1, so, $\aa \ra 4(D-1)^2 \bb $. As usually, we make a shift:
\be
\aa=4(D-1)^2\bb + \ep
\label{sh}
\ee
 It already shows that the big
weight $\bb^{Area}$ will be cancelled in modified SC expansion
(\ref{msurf}). Going to the local limit in (\ref{chese}) we obtain
\ba
  <\aa_{-n}> &    \\
& \simeq \Big([2D-2]^{-2n} \1-n {1\over 4(D-1)^2} <\ep > + ...\Big)  \\
& \simeq  \bb <\N F_{\mu\nu}^2(x)>_{QCD}> +\ \  \mbox{substractions} \\
\label{wcon}
\ea
where $F_{\mu\nu}=A_\nu,_\mu A_\mu,_\nu +[A_\mu, A_\nu]$ is the
gauge field strength.

We see that the weights of the modified SC
expansion are defined through the local condensates of
the gluon field $A_\mu(x)$ in QCD and can in
principle be calculated perturbatively, from the Feynman perturbation
technique.  All terms proportional to
the powers of cut-off in (\ref{wcon}) should be substracted,
 since they should cancel with
the contact terms, as happend for the vector-field.

However, the perturbative calculation cannot give us the most
interesting exponentially small terms in the gluon condensates, which
should in principle define the renormalized string tension. This problem
cannot be solved if we are unable to calculate the corresponding
sum over surfaces (\ref{msurf}) and get the effective action for the $\aa$
variables. The only purpose of the relations found here, between the
gluon condensates and the momenta of the lagrange multiplier, was to
show the existence of the new nontrivial weights in the modified SC
expansion leading to a lattice string representation for the
multicolour gauge theory.

\newsection{Conclusions}

Presenting this lattice construction of the random surface representation in
QCD, we did not want to create an impression that a real
chromodynamical string theory is around the corner.   All of this can
serve only as an intuitive picture from which one can probe
different continuum string world sheet lagrangians.
We tried here to demonstrate in  principle
the existence of the modified strong
coupling expansion with the weights of the expansion being defined
selfconsistently in the physical weak coupling phase. We managed to
demonstrate the existence of these weights in a toy one-plaquette
model, and then we gave some hints as to how to calculate them through
gluon condensates.

What are the main ingredients of this world sheet lagrangean?

As it was noted in papers \cite{Kostov} and \cite{Polyakov}, the fact
that the Wilson loop average should obey the Makeenko-Migdal equations
implies the existence (finiteness) of the area variational derivative of the
Wilson loop \cite{MigMak}. This means that the cosmological constant term
should be absent, and the action contains the term:
\be
\int d^2\xi T_{\mu \nu} \ep_{ab} \d_a X_\mu \d_b X_\nu
\label{xi}
\ee
where $X_\mu(\xi)$ is the world sheet coordinate in the physical space,
and $T_{\mu\nu}$ is either the independent antisymmetric field on the
world sheet (in \cite{Kostov}), or an antisymmetric
 function of $X_\mu$ (in \cite{Polyakov}).

This might be a good starting point but it is far from  the final
action. According to our construction of this paper, a probable object
to be added with a new coupling constant to this action would be a saddle like
configuration. This saddle is quite singular: for example, for the
saddle of the second order shown in fig.3, we have a deficit of the
angle (curvature) equal to $4\pi$ (since it consists from 4 plaquettes)
and concentrated in one point on the on the world sheet. The radii of
curvature at these points are equal to zero. Nevertheless, this singular
object is localized on the world sheet and can in principle be
described by a local operator in the action, depending on the extrinsic
geometry. The problem is to guess the continuous form of these terms for
saddles of any orders. Their coupling constants should be defined
from the unitarity condition, say in the form of the back-tracking
independence of the Wilson average (see section 3).

One possibility to describe the external geometry is to introduce the
external curvature-dependent terms, like
\be
\int d^2\xi \sqrt{g} [\Delta(g) X_\mu]^2
\label{excu}
\ee
as was proposed by Polyakov.
It is not clear whether this simple introduction of the extrinsic
geometry dependence is sufficient to describe the same effects as
saddles do for QCD.

The saddles seem to be   relevant factors for the whole problem.
In 2 dimensions, they define the so-called branch points of the
world sheet \cite{Kos2,Gross,Kostov,KazDou},  like on the Riemann
surface of the function $y=\sqrt{x}$.  This branch point can have a
negative weight, as seen from the expression for the Wilson average
with the loop C encircling twice some two dimensional domain of the
area A \cite{KazKos}:
\be
W(A)=e^{-\bb^{-1}A} (1- \bb^{-1}A)
\label{doubl}
\ee
The last term in the pre-exponent emerges from the entropy of the branch
point on the minimal area surface covering this contour. The weight of
the branch point is negative, and for $A>\bb$ the Wilson loop becomes
negative. One might wonder whether the same phenomenon could happen in
the four dimensional QCD in the deep confinement regime, when only the
minimal area gives the main contribution. For the same contour the
minimal surface should not look different as in two dimensions, with
the same branch point.

The only way to check it is by a Monte-Carlo simulation for the
lattice gauge theory. A good indication of the relevance of special saddle
points on the world sheet of the QCD string would be the negative
sign of this Wilson average. It would be a completely non-perturbative
result since in perturbation theory the Wilson average is always
close to one. Work on this calculation is in progress on the APE in
Rome \cite{APE}.

Of course, one can easily imagine a much more pessimistic scenario. The
saddles could be quite big and densely distributed on a typical world
sheet for the physically relevant values of $\bb$. Such a condensation
of saddles might completely cripple the surface, and the whole surface
description might be irrelevant in this situation. This looks to be a
probable scenario for big $\beta$, as we have already shown in the one
plaquette model, where big saddles dominated in the simple sum over
"surfaces". However, one might hope that these saddles are sufficiently
rare on the surface in the deep confinement regime. If the worst
scenario turns out to be true, it might mean that our hopes for a
string representation of QCD are hopeless.


\vspace{.8cm}

\begin {flushleft}
{\bf Acknowledgements}
The author would like to thank E.Brezin, I.Kostov, A.Polyakov and T.Wynter
for helpful discussions.
\end{flushleft}

\newpage
\begin {flushleft}
{\bf Figure Captions}

Fig.1. The typical strong coupling diagram for the vector field.

Fig.2. Glueing together of two plaquettes after a link  variable
integration.

Fig.3. Formation of saddles, tubes et.c. from four plaquettes after a
 link variable integration.

Fig.4 Multiple saddle.

Fig.5 Backtracking condition for the wilson loop, following from the
unitarity.

\end{flushleft}

\end{document}